\def\draftversion{false}
\RequirePackage{ifthen}
\ifthenelse{\equal{\draftversion}{true}}{
  \documentclass[citeautoscript,floatfix,aps,prl,twocolumn,preprintnumbers,
      superscriptaddress,longbibliography,amsmath,amssymb]{revtex4-1}
      \usepackage{showlabels}
}{
  \documentclass[citeautoscript,floatfix,aps,prl,twocolumn,
      superscriptaddress,longbibliography]{revtex4-1}     
}

\usepackage[utf8]{inputenc}
\usepackage{amsmath}
\usepackage{graphicx}
\usepackage{epstopdf}
\usepackage{natbib}
\usepackage{array}
\usepackage{dcolumn}% Align table columns on decimal point
\usepackage{bm}% bold math
\usepackage{siunitx} % ALIGING TABLE COLUMNS BY DECIMAL POINT
\usepackage{subfigure} % Using subfigures

\usepackage{soul}  % \st    for strike-out

\usepackage[usenames,dvipsnames]{color}

\soulregister\cite7

\ifthenelse{\equal{\draftversion}{true}}{
  \marginparwidth 2.7in
  \marginparsep 0.5in
  \newcounter{comm} % counter for commentaries
  \def\commnext{\stepcounter{comm}}
  \def\commtext{{\bf\color{blue}[\arabic{comm}]}}
  \def\commmar{{\bf\color{blue}[\arabic{comm}]}}
  \def\dvm#1{\commnext\marginpar{\small DV\commmar: #1}\commtext}
  \def\cdm#1{\commnext\marginpar{\small CED\commmar: #1}\commtext}
  \def\msm#1{\commnext\marginpar{\small MS\commmar: #1}\commtext}
  \def\asm#1{\commnext\marginpar{\small AS\commmar: #1}\commtext}
  \def\miq#1{\commnext\marginpar{\small MR\commmar: #1}\commtext}
  \def\mlab#1{\marginpar{\small\bf #1}}
  
}{
  \def\dvm#1{}
  \def\cdm#1{}
  \def\msm#1{}
  \def\asm#1{}
  \def\miq#1{}
  \def\mlab#1{}
  
}
\def\x{K}
\def\bx{\mathbf{K}}

\begin{document}

\title{Macroscopic polarization from nonlinear gradient couplings}

\author{Massimiliano Stengel}
\affiliation{Institut de Ci\`encia de Materials de Barcelona 
(ICMAB-CSIC), Campus UAB, 08193 Bellaterra, Spain}
\affiliation{ICREA - Instituci\'o Catalana de Recerca i Estudis Avan\c{c}ats, 08010 Barcelona, Spain}
\email{mstengel@icmab.es}

\date{\today}

\begin{abstract}
We show that a lattice mode of arbitrary symmetry induces a well-defined macroscopic 
polarization at first order in the momentum and second order in the amplitude.
We identify a symmetric flexoelectric-like contribution, which is sensitive
to both the electrical and mechanical boundary conditions, and an antisymmetric 
Dzialoshinskii-Moriya-like term, which is unaffected by either.
We develop the first-principles methodology to compute the 
relevant coupling tensors in an arbitrary crystal, which 
we illustrate with the example of the antiferrodistortive 
order parameter in SrTiO$_3$.
\end{abstract}

 %Lattice dynamics: first-principles theory
\maketitle

%%%%%%%%%%%%%%%%%%%%%%%%%%%%%%%%%%%%%%%%%%%%%%%%%%%%%%%%%%%%%%%%%%%%%%%%%%%%%%%%

The interaction between structural, polar and magnetic degrees of freedom in 
multiferroics has long been identified as a promising source of advanced material 
functionalities.
The recent focus on inhomogeneous structures such as skyrmions~\cite{Fert-17}, 
domain walls~\cite{Meier-15} and vortices~\cite{Behera-22,Naumov-04} has renewed 
the interest in 
the so-called \emph{Lifshitz invariants} (LIs), i.e., 
coupling terms that depend on the first gradient of one order parameter component.
LIs play a key role in the stabilization of
spatially modulated phases~\cite{Eliseev13,Kostya-22,Pottker16}
and often determine their emerging physical properties.
A paradigmatic example is the macroscopic 
Dzyaloshinskii-Moryia~\cite{Dz-58,Moriya-60} (DM) interaction,
\begin{equation}
\label{dmi}
E_{\rm DM} = \zeta {\bf P} \cdot \left[ \bm{\phi} (\bm{\nabla} \cdot \bm{\phi}) - (\bm{\phi} \cdot \bm{\nabla} )\bm{\phi} 
  \right].
\end{equation}
where $P$ is the macroscopic polarization, and $\bm{\phi}$
may correspond to the magnetic~\cite{Mostovoy06} 
or antiferromagnetic~\cite{Sparavigna94} degrees of freedom. 
[Realizations of Eq.~(\ref{dmi}) in other contexts, e.g., in 
liquid crystals~\cite{Sparavigna09} also exist.]
The importance of Eq.~(\ref{dmi}) lies in its topological character,~\cite{Xiao-09} 
and the rich phenomenology it can lead to, ranging from the switchable   
$P$ in ferroelectric multiferroics~\cite{Mostovoy06} to the stabilization of
incommensurate spin orders in broken-symmetry environments.

Another category of LIs involves flexoelectric-like terms, whereby $P$ is coupled to the gradient 
of a symmetric second-rank tensor field, ${\bf s}$,
\begin{equation}
\label{fxe}
E_{\rm flexo} = \frac{\x_{\alpha \beta \gamma \lambda}}{2}
 \left[\frac{\partial P_\alpha}{\partial r_\beta} s_{\gamma \lambda} -  P_\alpha \frac{\partial s_{\gamma \lambda}}{\partial r_\beta} \right].
\end{equation}
In the original form of Eq.~(\ref{fxe}), ${\bf s}$ corresponds to the elastic 
strain,~\cite{pavlo_review,Chapter-15} and the coupling is harmonic in 
the order parameter amplitudes.
More recently, Eq.~(\ref{fxe}) was generalized~\cite{Eliseev13} to a much broader 
class of \emph{nonlinear} couplings, where 
$s_{\gamma \lambda} = \phi_\gamma \phi_\lambda$ is the dyadic product of 
two (pseudo)vectors
[e.g., the ferroelectric polarization~\cite{Korzhenevskii-81} 
or the antiferrodistortive (AFD) tilts~\cite{Eliseev13} in perovskite-structure oxides].
Regardless of symmetry or the physical nature of ${\bf s}$, 
the coupling tensor $\x_{\alpha \beta \gamma \lambda}$ is a 
universal property of all crystals, hence its fundamental and practical 
interest.

Research efforts are currently directed at 
exploring practical realizations of these ideas in 
a variety of materials and order parameter types.~\cite{Pyatakov-15} 
It would be highly desirable, for instance,  
to find nonmagnetic analogues of Eq.~(\ref{dmi}),
in contexts where the strength of the coupling constant 
$\zeta$ is not limited by weak relativistic effects.~\cite{Zhao-21}
The so-called ferroelectric DM interaction,~\cite{Zhao-21,Chen-22}
which involves the polarization itself as the primary order parameter,
appears as an especially promising candidate. 
An antiferrodistortive (AFD) realization of Eq.~(\ref{dmi}) was also 
hinted at in Ref.~\cite{Schiaffino17}, although the relationship between
the ``rotopolar'' coupling described therein and Eq.~(\ref{dmi}) is not
immediately obvious.
Meanwhile, additional \emph{indirect} contributions to $P$ have also been 
pointed out, either involving the strain (``flexo-roto''~\cite{morozovska-12b} effect 
in the case of tilts) or other nonpolar degrees of freedom (e.g., the 
antiferroelectric $R$-mode of Refs.~\onlinecite{Schiaffino17} 
and~\onlinecite{Casals18}).
The coexistence of several effects, whose mutual relationship is
sometimes paradoxical,~\cite{Eliseev13} complicates the understanding of
flexo- and DM-type couplings, calling for a fundamental treatment.

The main priority for microscopic 
theory lies in clarifying the 
physical mechanisms that 
generate a polarization in inhomogeneous ferroic structures, either directly
via Eq.~(\ref{fxe}) and~(\ref{dmi}), or via the aforementioned indirect routes. 
In particular, it is of central importance to know
whether these effects are well-defined bulk properties 
of the crystal, or whether they are plagued by ambiguities (e.g., due
to boundary issues) as in the case of flexoelectricity.~\cite{artlin,Royo22} 
At the same time, it would be desirable to establish an efficient and
accurate methodological framework to predict the value of the relevant
coupling coefficients in real materials, e.g., via first-principles 
techniques.
Selected components of the rotopolar tensor in SrTiO$_3$ 
have been calculated already;~\cite{Schiaffino17} 
however, conceptual and technical difficulties 
with the treatment of spatial dispersion effects at nonlinear order have 
so far thwarted the development of a full-fledged theory.

Here we provide a unified first-principles theory of both 
flexo- 
and DM-type couplings
by expressing them as, respectively, the 
symmetric and antisymmetric parts of the same fourth-rank tensor.
Based on this result, we argue that an arbitrary inhomogeneous 
field $\bm{\phi}$ always couples to polar degrees of freedom
via both mechanisms, with the special case where 
${\bf P}$ and $\bm{\phi}$ are \emph{the same mode} as an 
interesting exception.
We further show that the DM-like 
coefficient $\zeta$ is a well-defined physical property of the crystal, 
while the flexo-type 
tensor, $\x_{\alpha \beta \gamma \lambda}$, is not. 
The reason lies in the macroscopic elastic and electrostatic interactions,
which contribute to the latter but not to the former. 
Similarly to the flexoelectric case, these long-ranged (``nonanalytic'', in the language of 
perturbation theory) terms 
lead to ambiguities in the definition of the reference 
electrostatic potential and the center of 
mass of the cell,~\cite{artlin,Royo22} which must be 
adequately treated to guarantee the internal consistency of 
the theory.~\cite{Dieguez-22} 
From a practical point of view we recast
the nonlinear interaction between 
modulated order parameters as well-defined third derivatives of the total energy.
The long-wavelength expansion~\cite{Royo19,Royo22} of the latter, 
which we treat in the framework of density-functional perturbation 
theory~\cite{Baroni-01,Gonze-97} (DFPT), readily yields the coupling 
constants of Eq.~(\ref{fxe}) and Eq.~(\ref{dmi}) at first order in the momentum.
Calculations are performed with minimal effort via 
the recently developed~\cite{Royo19,Royo22} long-wave module of {\sc Abinit},~\cite{abinit,Romero-20} 
in combination with a post-processing tool that we have implemented and
tested as part of this work.
As a numerical demonstration, we focus on the leading terms involving the AFD 
order parameter in SrTiO$_3$.

Following Ref.~\onlinecite{Schiaffino17}, we base our derivations on 
\emph{unsymmetrized} inhomogeneous couplings of the type
\begin{equation}
\label{uns}
E_{\rm uns} = -W_{\alpha \beta \gamma \lambda}
 p_\alpha \frac{\partial \phi_\gamma }{\partial r_\beta} \phi_\lambda,
\end{equation}
where $E_{\rm uns}$ is an energy (per primitive 5-atom cell), 
$\bm{\phi}({\bf r})$ is the main order parameter, 
and the field ${\bf p}({\bf r})$ corresponds to some polar lattice mode of the 
crystal.
Eq.~(\ref{uns}) is the most general trilinear coupling between $\bm{\phi}({\bf r})$
and ${\bf p}({\bf r})$ occurring at first order in the gradient expansion; any other
expression can be written as a linear combination thereof.
To verify this point explicitly in the case of 
Eq.~(\ref{fxe}) and~(\ref{dmi}), 
it suffices to separate the symmetric and antisymmetric 
contributions with respect to the last two indices,
$W_{\alpha \beta \gamma \lambda} = W_{\alpha \beta (\gamma \lambda)} + W_{\alpha \beta [\gamma \lambda]}$.
Within the assumed cubic symmetry, elementary calculus leads then to 
\begin{equation}
\label{wxz}
W_{\alpha \beta \gamma \lambda} = \underbrace{2\x_{\alpha \beta \gamma \lambda}}_{W_{\alpha \beta (\gamma \lambda)}} + 
 \underbrace { \zeta (\delta_{\alpha \gamma}\delta_{\beta \lambda} -
              \delta_{\alpha \lambda}\delta_{\beta \gamma})
              }_{W_{\alpha \beta [\gamma \lambda]}},
\end{equation}
which establishes the formal link between Eq.~(\ref{uns}) and Eqs.~(\ref{fxe}--\ref{dmi}).
In a cubic crystal, $\bx$ has three independent entries ($\x_{11}=\x_{1111}$,
$\x_{12}=\x_{1122}$ and $\x_{44}=\x_{1212}$), similarly to
the flexoelectric tensor. %(Following the established conventions, we shall refer to 
%them as $\x_{11}=\x_{1111}$, $\x_{12}=\x_{1122}$ and $\x_{44}=\x_{1212}$.) 
These, in combination with the DM-type scalar $\zeta$, account
for the four components of the tensor ${\bf W}$; the latter coincides with the ``rotopolar'' 
coupling of Ref.~\onlinecite{Schiaffino17} %if $\bm{\phi}$ 
in the AFD case.

The special case where $\bm{\phi}={\bf p}$, of relevance to the recently proposed 
``electric DM interaction''~\cite{Zhao-21,Chen-22}, deserves a separate discussion.
Eq.~(\ref{uns}) reduces then to 
Eq.~(\ref{fxe}) via a permutation
of indices and integration by parts.
This means that the DM-type coupling of Eq.~(\ref{dmi}) is redundant in
this case: the flexo-type expression of Eq.~(\ref{fxe}) describes
the trilinear self-interaction of a polar vector field in full generality.
Assuming cubic symmetry of the undistorted crystal, Eq.~(\ref{fxe})
adopts the following compact form,
\begin{equation}
\label{pinv}
E_{\rm p} = \x p^2 \bm{\nabla} \cdot {\bf p}.
\end{equation}
where $\x = \x_{12}-\x_{44}$ is a single material coefficient, 
and $p^2={\bf p} \cdot {\bf p}$. 
The remaining independent 
components of the $\bx$-tensor ($\x_{11}$ and $\x_{12}+\x_{44}$) are 
irrelevant at the bulk level as they do not contribute to the forces nor to the
energy.
Crucially, Eq.~(\ref{pinv}) depends directly on the longitudinal 
components of ${\bf p}$, which are typically suppressed by depolarizing
effects; for this reason, henceforth we shall restrict to our attention to 
cases where the primary order parameter $\bm{\phi}$ is nonpolar.

To work our way towards a first-principles expression, we need to specify the 
microscopic nature of the field variables entering Eqs.~(\ref{uns}).
Following Ref.~\onlinecite{Dieguez-22}, we use a perturbative approach
in terms of monochromatic lattice distortions of the type
\begin{equation}
u^l_{\kappa \alpha} = u^{\bf q}_{\kappa \alpha} e^{i{\bf q}\cdot{\bf R}^{(0)}_{l\kappa}}.
\label{mapping}
\end{equation}
Here $\kappa$ and $l$ are sublattice and cell indices, respectively; $u^l_{\kappa \alpha}$ 
indicates the atomic displacement along the Cartesian direction $\alpha$; 
${\bf R}^{(0)}_{l\kappa}$ stands for the unperturbed atomic locations 
in the high-symmetry reference structure; ${\bf q}$ is the momentum.
The 
microscopic representation of the continuum fields is then defined as 
\begin{equation}
u^{{\bf q}}_{\kappa \alpha} = \langle \kappa \alpha| p_\beta \rangle  p_\beta^{\bf q} + \langle \kappa \alpha| \phi_\beta \rangle  \phi_\beta^{\bf q},
\end{equation}
where the symbol $\langle \kappa \alpha| v \rangle$ corresponds~\cite{sm}
to the eigendisplacements of a given phonon mode $| v \rangle$,
and ${\bf v}^{\bf q}$
refers to the Fourier representation of the field ${\bf v}({\bf r})$.
(Bra and kets refer to real vectors in the $3N$-dimensional space of 
the atomic displacements, where $N$ is the number of basis atoms in the cell.~\cite{sm})

Based on the above, we can express Eq.~(\ref{uns}) in reciprocal space 
as a three-phonon vertex,
\begin{equation}
E_{\rm uns} = -i q_\beta W_{\alpha \beta \gamma \lambda} p_\alpha^{{\bf -q} {\bf -q}'}  
    \phi_\gamma^{\bf q} \phi_\lambda^{\bf q'}.
\end{equation}
In the ${\bf q,q'} \rightarrow 0$ limit, we can then write the tensor 
${\bf W}$ in terms of the third derivatives of the total energy $E$, 
\begin{equation}
\label{d3e}
\frac{\partial^3 E}{\partial p_\alpha^{\bf -q} \partial \phi_\gamma^{\bf q} \partial \phi_\lambda^0}
= \langle p_\alpha | \frac{\partial \Phi^{\bf q}}{\partial \phi^0_\lambda} | \phi_\gamma \rangle.
%-\frac{\partial^2 f_\alpha^{\bf q}}{\partial \phi_\gamma^{\bf q} \partial \phi_\lambda^0} \simeq
% -i q_\beta \Omega W_{\alpha \beta \gamma \lambda},
\end{equation}
or equivalently as the first derivative of the force-constant matrix, $\Phi^{\bf q}$,
with respect to the homogeneous perturbation $\phi^0_\lambda$.
By recalling~\cite{artlin,Royo22} the long-wave expansion of $\Phi^{\bf q}$,
%\begin{equation}
$\Phi^{\bf q} \simeq \Phi^{(0)} - i q_\beta \Phi^{(1,\beta)}$,
%\end{equation}
we arrive then at a closed expression for the ${\bf W}$-tensor components as
projection on the polar mode $\langle p_\alpha |$ of the \emph{force-response tensor}
$|w_{\beta \gamma \lambda} \rangle$,
\begin{equation}
\label{walp}
W_{\alpha \beta \gamma \lambda} = %\frac{1}{\Omega} 
   \langle p_\alpha | w_{\beta \gamma \lambda} \rangle, \quad
   |w_{\beta \gamma \lambda} \rangle = \frac{\partial \Phi^{(1,\beta)}}{\partial \phi^0_\lambda} | \phi_\gamma \rangle.
%   \underbrace{  \frac{\partial \Phi^{(1,\beta)}}{\partial \phi^0_\lambda} | \phi_\gamma \rangle
%       }_{|w_{\beta \gamma \lambda} \rangle}.
\end{equation}
Thanks to cubic symmetry, Eq.~(\ref{walp}) allows one to capture all the independent components 
of ${\bf W}$ at once as part of a single linear-response calculation; the flexo-like and
DM-like contributions are then extracted via Eq.~(\ref{wxz}).
%
%(Of course, Eq.~(\ref{walp}) is specific to the polar mode $|p_\alpha\rangle$; 
(Whenever appropriate, the mode index will be indicated with a superscript,
either in the form $W^{(i)}_{\alpha\beta\gamma\lambda}$ or $W^{[i]}_{\alpha\beta\gamma\lambda}$
for the normal-mode or symmetry-adapted~\cite{Hong-13} sublattice representation~\cite{sm} 
of the tensors, respectively.)

Our next goal is to understand whether ${\bf W}$ (or its decomposition
into $\bx$ and $\zeta$) 
is a well-defined physical property of the crystal.
A first concern lies in the definition of the force-response tensor, 
$|w_{\beta\gamma\delta}\rangle$, via a long-wave expansion of $\Phi^{\bf q}$.
To perform the latter operation, short-circuit
electrical boundary conditions need to be imposed,~\cite{artlin} which implies 
setting to zero the macroscopic electrostatic potential, $V^{\rm mac}$, in
the calculations.
$V^{\rm mac}$ is, however, ill-defined in a periodic crystal,~\cite{resta-dp} 
which leads to a \emph{reference potential ambiguity} in the definition of
the flexo-type coefficients.~\cite{artlin,Stengel-16,Royo-22}
As this issue only affects the longitudinal components of the polarization 
response, which are expected to be small, we won't delve into it further here;
%and 
%leave a detailed discussion to a future work. 
in any case, the DM-type %coefficients $W_{\alpha \beta [\gamma\delta]}$, and
%hence the 
constant $\zeta$ is manifestly 
unaffected by electrostatics, 
due to the transverse nature of Eq.~(\ref{dmi}).

A second issue %source of trouble 
concerns the translational
freedom of the polar mode eigendisplacement vector, which
is only defined modulo a rigid shift of the cell.~\cite{Dieguez-22}
%imposition of short-circuit electrical boundary conditions that is 
%implicit in the definition of W as a long-wave expansion coefficient.
%
Based on the criteria of Ref.~\onlinecite{Dieguez-22}, a necessary condition 
for a material property to be ``well defined'' is its invariance 
%it needs to be invariant  
%we require the definition of ${\bf W}$ to be invariant 
with respect to the following transformation,
\begin{equation}
\label{transformp}
|p_\alpha'\rangle = |p_\alpha\rangle + \lambda, % |u_\alpha\rangle,
\end{equation}
where %$|u_\alpha\rangle$ is the \emph{acoustic} eigendisplacement vector and 
$\lambda$ an arbitrary constant.
%
%(Due to translational invariance, the polar distortion is only 
%defined modulo a rigid shift of the cell.~\cite{Dieguez-22})
%
To understand the impact of Eq.~(\ref{transformp}) on ${\bf W}$,
recall that the acoustic eigendisplacement vector 
reduces to a translation~\cite{Stengel-16,Dieguez-22} 
regardless of the microscopics, $\langle \kappa \alpha| u_\beta \rangle = \delta_{\alpha \beta}$.
This implies that 
\begin{equation}
{W}'_{\alpha\beta\gamma\delta} = {W}_{\alpha\beta\gamma\delta} + %\frac{\lambda}{\Omega} 
  \lambda \langle u_\alpha |w_{\beta\gamma\delta} \rangle,
\end{equation}
where $\langle u_\alpha|w_{\beta\gamma\delta} \rangle$ is a net elastic force 
on the cell as a whole that %is associated with $\phi_\lambda \partial \phi_\gamma /\partial \phi_\beta$.
arises in a locally inhomogeneous order parameter $\bm{\phi}$.
That such a force does not vanish is a direct consequence of the \emph{strain coupling}
\begin{equation}
\label{rotos}
E_{\rm sc} = -R_{\alpha\beta\gamma\delta} \varepsilon_{\alpha\beta} %\frac{\partial u_\alpha}{\partial r_\beta} 
\phi_\gamma \phi_\delta,
\end{equation}
which is always allowed by symmetry. Since the force is the 
divergence of the stress, a trivial integration by parts leads to
the following \emph{sum rule},
%we first establish the following \emph{sum rule},
%By plugging Eq.~(\ref{transformp}) into Eq.~(\ref{walp}), we readily 
%obtain
\begin{equation}
\label{sumrule}
%W'_{\alpha \beta \gamma \lambda} = W_{\alpha \beta \gamma \lambda} + \lambda
%-\frac{1}{2\Omega} 
-\frac{1}{2} \sum_\kappa \langle \kappa \alpha |w_{\beta \gamma \lambda} \rangle = R_{\alpha\beta\gamma\lambda},
\end{equation}
relating the sublattice sum of the force-response tensor $|w_{\beta \gamma \lambda} \rangle$
to the \emph{strain coupling} tensor, ${\bf R}$.
After observing that $R_{\alpha \beta\gamma\lambda}$ is symmetric both in $\alpha \beta$ and
$\gamma\lambda$, we arrive at the following transformation law for the
coupling coefficients,
\begin{equation}
\label{transformxi}
\x_{\alpha\beta\gamma\delta}' = \x_{\alpha\beta\gamma\delta} - \lambda R_{\alpha\beta\gamma\delta}, \qquad \zeta' = \zeta.
\end{equation}
Eq.~(\ref{transformxi}) is one of the central results 
of this work, showing that the DM-like coupling constant, unlike $\bx$, %transforms as $\zeta' = \zeta$, i.e., 
%ones, $\nu^{\rm lm}_{\alpha \beta [\gamma \lambda]}$,
%are 
is indeed invariant with respect to Eq.~(\ref{transformp}), and hence
a well-defined bulk property, as anticipated earlier.

Notwithstanding the aforementioned ambiguity of $\bx$, %we shall now demonstrate that 
the information contained in it is crucial to obtaining a % yields a  %the force-response tensor $|\x\rangle$ yields a
well-defined value of the local 
polarization at leading order in $\phi$ and ${\bf q}$.
To see this, we assume in the following that the fields are modulated along a single
direction $\hat{s}$ and constant along the normal planes. (This is appropriate,
for example, to modeling a domain wall oriented along ${\hat{s}}$.)
Within these mechanical boundary conditions, we obtain (see Section S8~\cite{sm}) the 
relaxed electrical polarization as (summation over repeated indices is implied)
\begin{equation}
\label{dist}
%p^{[i]}_\alpha 
P_\alpha = \frac{1}{\Omega} Z^{[i]}
%|\tau_\kappa\rangle = 
 {\Phi}_{ij}^{+} \left(
\tilde{\x}^{[j]}_{\alpha \gamma \lambda}(\hat{s}) \mathcal{S}_{\gamma\lambda,s}
 + \zeta^{[j]} \mathcal{A}_\alpha 
\right),
%\frac{\partial }{\partial s} (\phi_\gamma \phi_\lambda),
\end{equation}
where ${\Phi}^{+} $ is the pseudoinverse~\cite{born/huang,Hong-13,Royo-22} of the zone-center force-constants matrix;
$\mathcal{S}_{\gamma\lambda,s}=\partial (\phi_\gamma\phi_\lambda)/\partial s$ and 
$\mathcal{A}_\alpha = \phi_s \partial \phi_\alpha/\partial s -  \phi_\alpha \partial \phi_s/\partial s$
are the relevant symmetric and antisymmetric components of the nonlinear $\phi$-gradient tensor;
$Z^{[i]}$ are the mode dynamical charges;
and the renormalized flexo-like coefficients are
\begin{equation}
\label{tilxi}
\tilde{\x}^{[j]}_{\alpha \gamma \lambda}(\hat{s}) = \x^{[j]}_{\alpha \hat{s} \gamma \lambda} + 
 C^{[j]}_{\alpha \hat{s} \beta\hat{s}}
  [\mathcal{C}(\hat{s})]^{-1}_{\beta \sigma} R_{\sigma \hat{s} \gamma \lambda}.
%|w_{\gamma \lambda}(\hat{s}) \rangle = |w_{\hat{s} \gamma \lambda} \rangle + 2 |C_{\hat{s} \beta \hat{s}} \rangle 
%  [\mathcal{C}(\hat{s})]^{-1}_{\beta \sigma} R_{\sigma \hat{s} \gamma \lambda}.
\end{equation}
Here $C^{[j]}_{\alpha \hat{s} \beta \hat{s}}$ and 
$\mathcal{C}_{\beta \sigma}(\hat{s}) = \mathcal{C}_{\beta \hat{s} \sigma \hat{s}}$ are the 
projections along $\hat{s}$ of the flexoelectric coupling~\cite{sm} and elastic tensors, respectively.
The second term in Eq.~(\ref{tilxi}) originates from the relaxation of the acoustic modes, 
which produce a strain gradient (and hence atomic forces via flexoelectricity) 
at first order in ${q}$.

By combining the sum rule Eq.~(\ref{sumrule}) with its flexoelectric
counterpart,~\cite{artlin,Royo-22} $\sum_j C^{[j]}_{\alpha \beta \gamma \lambda} = \mathcal{C}_{\alpha \beta \gamma \lambda}$, 
it is straightforward to verify that the sublattice sum of the renormalized 
force-response coefficients, $|\tilde{\x}_{\gamma \lambda}(\hat{s})\rangle$, 
identically vanishes. This guarantees~\cite{born/huang} that the 
total polarization $P_\alpha$ is well defined, proving our point. 
Conversely, the individual contributions to $P_\alpha$ associated with
the two terms in Eq.~(\ref{tilxi}) depend on how the pseudoinverse
is constructed,~\cite{Dieguez-22} and are therefore ill defined
as stand-alone properties.
%Eq.~(\ref{dist}) yields well-defined
%values (modulo a trivial global shift) of the polar distortions $\tau_\kappa$.
%
Such intimate relationship between the 
direct flexo-like contribution to the atomic forces [first term in 
Eq.~(\ref{tilxi})] and the \emph{nonanalytic elastic contribution} 
of the second term provide a nice illustration of the 
covariance principle of Ref.~\cite{Dieguez-22}, which we
generalize here to the nonlinear regime.
These conclusions have direct implications for the continuum modeling
of inhomogeneous ferroelectric~\cite{Yudin-12,Gu-14,Chen-22} and
ferroelastic~\cite{Eliseev13,morozovska-12b,Schiaffino17,Schranz-20,Troster-22}
structures, where the aforementioned two mechanisms play a central role.

As a representative demonstration of the above arguments, 
we consider the case where the field $\bm{\phi}$ corresponds 
to the out-of-phase antiferrodistortive (AFD) tilts in perovskite-structure oxides,
with SrTiO$_3$~\cite{Schiaffino17} % and PbZrO$_3$~\cite{Kostya-22} as 
as testcase.
Calculations of the rotopolar~\cite{Schiaffino17} force-response tensor $|w_{\beta \gamma \lambda}\rangle$, 
(its symmetric part, 
$|\x_{\beta \gamma \lambda}\rangle = (|w_{\beta \gamma \lambda}\rangle + |w_{\beta \lambda \gamma }\rangle)/4$,
corresponds to the ``flexo-AFD'' effect described in Ref.~\cite{Eliseev13})
are carried out 
within the framework of the local-density approximation (LDA) to density-funcional theory
as implemented in the {\sc Abinit}~\cite{Gonze2016,Romero-20,hamann-13,pseudodojo} package. 
We use a nonprimitive cell of 10 atoms in order to accommodate a small uniform tilt $\phi^0_\alpha$ 
in the structure, which allows us to treat the third derivatives of Eq.~(\ref{walp}) via 
finite differences in $\phi^0_\alpha$.
(The parametrization of the AFD mode amplitudes, in length units, 
follows the established convention.~\cite{Cao-90,Schiaffino17}; relaxation of the 
antiferroelectric $R$-mode of Ti~\cite{Schiaffino17,Casals18} is fully accounted for
in the calculated $W^{[i]}_{\alpha\beta\gamma\lambda}$ coefficients.)
Numerical results,
details of the method and additional supporting data
are reported in Ref.~\cite{sm};
of particular note, we provide~\cite{sm} a stringent numerical 
proof of the sum rule, Eq.~(\ref{sumrule}), which we base 
on an independent calculation of the ${\bf R}$ 
(rotostriction~\cite{Eliseev13,morozovska-12b,Schiaffino17}) tensor.

\begin{figure}
\begin{center}
\includegraphics[width=3.2in]{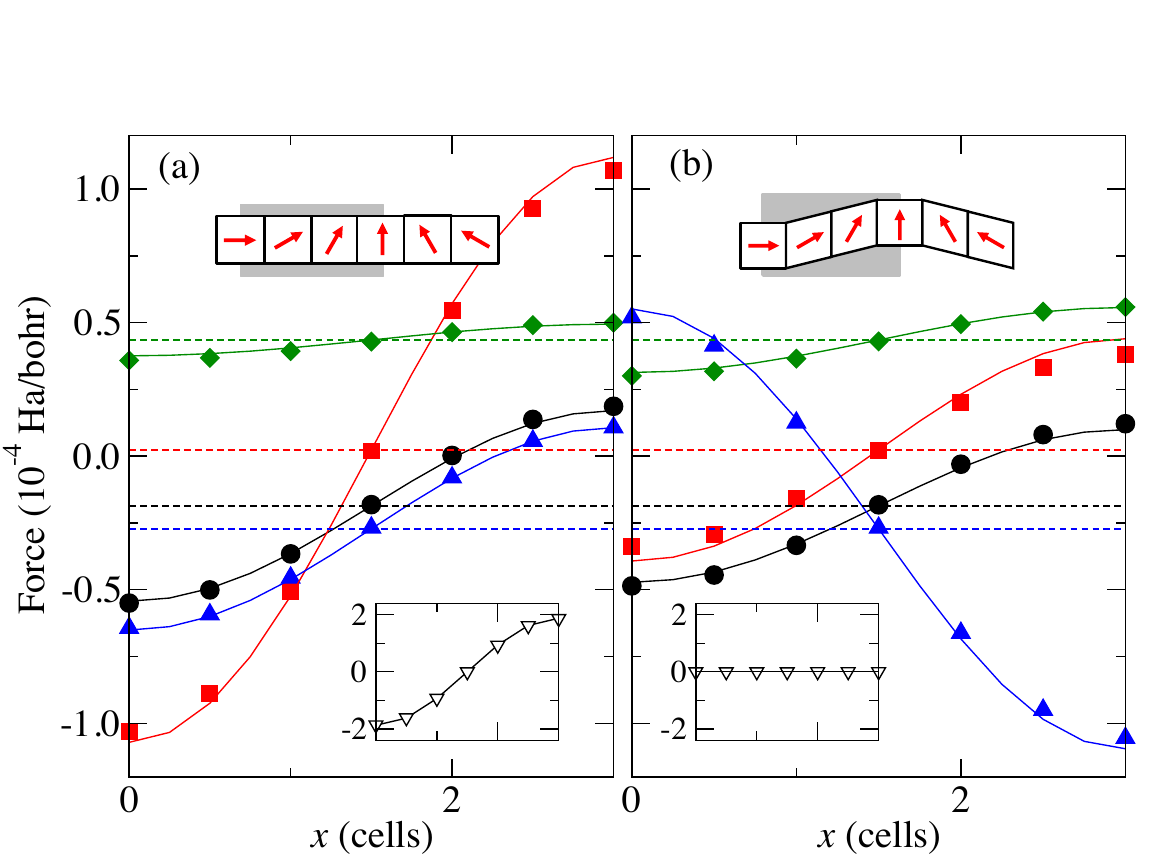}
\caption{\label{fig1} Comparison between the atomic forces extracted from a 
direct calculation of an AFD cycloid (symbols) and the predictions of the macroscopic 
model (solid curves). Forces on Sr (black circles), Ti (red squares), and the two 
$T_{1u}$ oxygen modes $\xi_3$ (blue triangles)
and $\xi_4$ (green diamonds)  are shown. Dashed lines show the uniform DM-like forces. 
(a): no elastic relaxation; (b): mechanical equilibrium. 
Lower insets show the sublattice sum of the forces: first-principles (triangles) and model (solid curve).
Upper insets schematically illustrate the relevant portion of the (fixed- or relaxed-strain) cycloidal pattern.} 
\end{center}
\end{figure}

To illustrate the physical meaning of the calculated 
rotopolar coefficients,
and as a further numerical validation thereof, we next consider a frozen-in
cycloidal~\cite{Schiaffino17} tilt pattern in the form %$\bm{\phi}({\bf r}) = \phi[\cos({\bf q\cdot r}),\sin({\bf q\cdot r}),0]$,
$\phi_s = \phi \cos({\bf q\cdot r})$, $\phi_r = \phi \sin({\bf q\cdot r})$, $\phi_z=0$,
where both the AFD pseudovector $\bm{\phi}$ and the propagation direction 
${\bf q}= q\hat{s}=q [\cos(\theta),\sin(\theta),0]$ lie in the pseudocubic $xy$ plane.
[Here $\hat{r}=\hat{z}\times \hat{s}$ is the in-plane direction that is orthogonal to ${\bf q}$.]
Our long-wavelength approach predicts, for the symmetry-adapted 
sublattice mode $[i]$, % an atom $\kappa$ located at 
%a given point ${\bf r}$ in the crystal, 
a \emph{geometric} force (both 
DM- and flexo-type couplings are linear in $P$, which implies an
improper~\cite{improper,bousquet/ghosez:2008,Fennie-05,Benedek-11} 
mechanism for local inversion-symmetry breaking) at a given point ${\bf r}$ in the 
crystal, whose transverse component reads as
%along $\hat{r}=\hat{z}\times \hat{q}$
%($\hat{z}$ spans the normal to the $xy$ plane)
\begin{equation}
\label{predict}
%\hat{r} \cdot {\bf f}_{\kappa}({\bf r}) 
f_r^{[i]}({\bf r})  = \phi^2 q \left[ \zeta^{[i]} + 
                               2 %\rangle \hat{r}\kappa |\x_{{\hat s}{\hat r}{\hat s}}\rangle 
                                \x^{[i]}(\hat{q}) \cos(2 {\bf q\cdot r})  \right].
\end{equation}
%Here $\hat{r}=\hat{z}\times \hat{q}$ is the in-plane direction that is orthogonal to ${\bf q}$; 
%
%$\zeta^{[i]}$ stand for the sublattice components of the isotropic DM-type force-response vector $|\zeta\rangle$,
%and 
($\x^{[i]}(\hat{q}) = \x^{[i]}_{rsrs}$
%\langle r \kappa | \x_{qrq}\rangle$ 
stand for
the 1212 components of the flexo-type tensor in the rotated ${\hat s},{\hat r},\hat{z}$
system.)
In Fig.~\ref{fig1}(a) we compare the prediction of Eq.~(\ref{predict}) with the
forces that we obtain via a direct first-principles calculation of the AFD cycloid.
(We use $\theta=\pi/4$, corresponding to  $\hat{s}\parallel [110]$ in the pseudocubic 
system, and $q=2\pi/(12 \sqrt{2} a_0)$, which we accommodate in a 120-atom supercell;
the tilt amplitude is set to $|\bm{\phi}|=0.02 a_0$, i.e., to a tilt angle of $2.3^\circ$.)
The agreement is excellent, with a discrepancy of the order of few percents at most.
Note the qualitative difference between the uniform DM-like contribution to $f_r^{[i]}$ (dashed lines), %${\bf f}_{\kappa}$,
%consistent with the topological character of Eq.~(\ref{dmi}), 
and the spatially modulated flexo-like 
term, which averages to zero in any periodic tilt pattern.

The uniform DM-like forces sum up to zero, consistent with the translational symmetry 
of the crystal and with our formal results.
Conversely, the flexo-like forces display the expected drift, shown in the inset of Fig.~\ref{fig1}(a), 
that originates from the strain coupling via Eq.~(\ref{sumrule}).
To verify that the net drift disappears at mechanical equilibrium, we determine the  
elastic displacement amplitude via $u(s) = -\phi^2R_{rsrs} / (2q \mathcal{C}_{rsrs}) \cos(2 {qs})$. 
%where $\mathcal{C}'_{66}$ and $R'_{66}$ are the relevant elastic and rotostriction constant
%in the rotated system. 
%
After incorporating $u(s)$ into the simulation cell, we 
recalculate the forces from first principles, and compare
them in Fig.~\ref{fig1}(b) with the predictions of
Eq.~(\ref{tilxi}). [The latter implies an additional contribution
to Eq.~(\ref{predict}) of the type $\Delta f^{[i]}_{r} = -4q^2 C^{[i]}_{rsrs} u(s)$.]
%where $C_\kappa({\hat q}) = \langle \kappa r| C_{qrq}\rangle$ is the relevant component of
%the flexoelectric force-response tensor.] 
Again, the agreement is excellent, and the 
elastic forces (inset) now vanish as expected.
%with the predictions of our macroscopic theory, which we derive from the 
%renormalized coefficients of 
%between the direct first-principles calculation of $\tilde{f}_\kappa$, which now
%sum to zero everywhere (see inset), and the predictions of Eq.~(\ref{tilxi}).
%
%In addition to validating our claims numerically, 

In addition to validating our claims numerically, this test clarifies 
the relation between the ``flexoantiferrodistortive''~\cite{Eliseev13} 
and ``flexo-roto''~\cite{morozovska-12b} effects (corresponding to
the second term in Eq.~(\ref{predict}) and to $\Delta f^{[i]}_{r}$, respectively) 
described in the recent literature, and the necessity to account for both in order to 
obtain quantitatively accurate physical answers.
Fig.~\ref{fig1} also 
%the results presented here 
demonstrates (once more~\cite{Dieguez-22}) the ability of our 
% macroscopic theory
\emph{second-principles}~\cite{Ghosez-22} macroscopic theory 
to predict the atomic forces (and hence the relaxed positions) in arbitrary inhomogeneous 
ferroic structures, with an accuracy that is comparable to that of direct \emph{ab initio} 
calculations.
This provides an ideal framework for multiscale simulations, 
as the coefficients of the  continuum Hamiltonian can be 
systematically validated against the microscopic reference in an exact limit.

The usefulness of the present theory is especially manifest in materials like 
SrTiO$_3$, where the lowest-frequency
(``soft'') mode at the zone center carries 99\% of the polar response to a 
static perturbation. This observation leads to %allows for 
a compact representation of the 
physical effects described insofar in terms of few materials-specific 
parameters; their calculated values are reported in Table~S9~\cite{sm}.
The information therein allows one to calculate the local polarization in 
an arbitrary inhomogeneous tilt structure via Eq.~(\ref{dist}).
For example, in the cycloid model of Fig.~\ref{fig1}, which is 
representative of the typical inhomogeneities of the order parameter in 
the ferroelastic phase of SrTiO$_3$, we obtain a macroscopic polarization
of $P=-0.9$ $\mu$C/cm$^2$. %which is significant. 
%
%Note that in our LDA calculations we obtain a relatively high soft-mode frequency 
%of $\omega=50.3$ cm$^{-1}$, corresponding to a static dielectric constant 
%of $\epsilon\simeq 900$. 
The effect increases linearly with the static dielectric constant,
$\epsilon$, which should lead to values of $P$ that are at least an order of 
magnitude larger at low temperature. 
($\epsilon\simeq 900$ in our LDA calculations.)
%$\xi=-0.013$ and $\zeta=-0.045$,
%$\phi^2 q = 0.00107725$, $Z=21.8$, $\Omega=7.283^3$ and $\chi^{-1}=0.0176$;
%this yields a macroscopic polarization of $P=-.000155$ (about 1 uC/cm2?)
%/Users/mstengel/Dropbox/Work/STO_validate/Cycloid/Validation/12b_new/README
%
% (the result is essentially exact 
%at the leading order in ${\bf q}$ and tilt amplitude and in absence of 
%depolarizing effects); 
%

In summary, %a broader perspective, 
the trilinear % couplings, together with the DM-like interaction 
Lifshitz invariants of Eq.~(\ref{dmi}) and~(\ref{fxe}) 
emerge here as a rich and barely tapped playground of 
potentially useful crystal properties, thus opening many opportunities 
for future research. 
We shall explore these promising  directions in the context of forthcoming publications.

\begin{acknowledgments}
        We acknowledge support from Ministerio de Ciencia
        Y Innovaci\'on (MICINN-Spain) through
        Grant No. PID2019-108573GB-C22;
        from Severo Ochoa FUNFUTURE center of excellence (CEX2019-000917-S);
        from Generalitat de Catalunya (Grant No. 2021 SGR 01519); and from
        the European Research Council (ERC) under the European Union's
        Horizon 2020 research and innovation program (Grant
        Agreement No. 724529).
\end{acknowledgments}

%%%%%%%%%%%%%%%%%%%%%%%%%%%%%%%%%%%%%%%%%%%%%%%%%%%%%%%%%%%%%%%%%%%%%%%%%%%%%%%%
\bibliography{merged,PZO}

\end{document}